# Option-Driven Design: Context, Tradeoffs, and Considerations for Accessibility

FRANK ELAVSKY, Carnegie Mellon University, fje@cmu.edu

In this micro-paper I outline the context and working definition for *option-driven design*, followed by several design negotiations, tradeoffs, and suggestions worth considering when choosing an option-driven design approach.

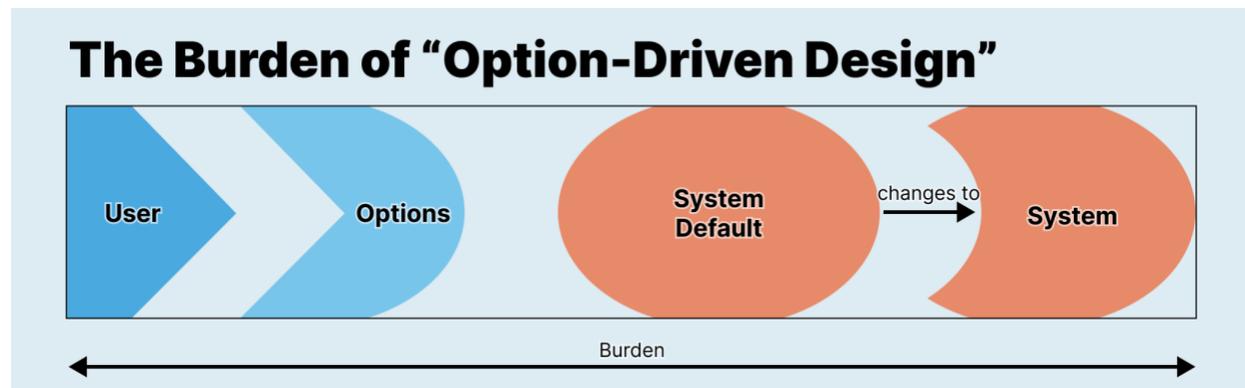

**Fig 1:** In "Option-Driven Design," users must interact with options and settings for systems to adapt to their needs. This approach places the burden on both the user and the system to make the interaction *between* user and system fit. The user must know and find which options they need and then adjust them. In addition, the system must be capable of robust change, similar to system change in *ability-based design.*

## 1. What is Option-Driven Design (ODD)?

Option-driven design is a design strategy to present users with options for manipulating the default logic and presentation of an interface system, such as through the adjustment of settings and preferences. Generally, when a user changes system state in this way, their chosen options or preferences are not overridden and will persist throughout other interactions (and even be inherited by other systems). ODD is often used in the context of software applications such as games, desktop and mobile operating systems, and internet browsers.

While ODD as a practice has been ubiquitous in computing systems for decades, an examination of ODD from the perspective of accessibility and its impacts is overdue. The immediate intention of this paper is to provide clarity to designers who are considering whether, why, and how to use ODD in a system. The broader intention of this paper is to stimulate new research, conversations, technical solutions, and ideas related to options and accessibility.

## 2. What is the context of accessibility & Option-Driven Design?

In contemporary accessible computing practices, designers and developers navigate complex design tensions, building systems that are assumed to fit to different user abilities. This practice of a designer engaging their own (and a system's) "ability assumptions" in regard to accessibility comes from the existing work of Wobbrock et al [14]. In the design space around ability assumptions, a designer recognizes that a system may have been built with assumptions about a user's abilities (such as the user's sight, motor functions, and more). For example, a trackpad or mouse is built with the assumption that the input it receives is from a user's hand and fingers, which are assumed to operate in the same way according to normative expectations that the designer has. The designer assumes all users have hands and fingers, all of which also operate in the same way.

These assumptions create problems for users (see Fig 2). When users must adapt to a system's assumptions, the burden of interaction is placed on the user. Users compensate in a variety of ways to fill the gaps left by a designer's assumptions, using different body parts, augmentations, or behaviors than the designer expected. Wobbrock et al propose that systems should recognize and adapt to a user's abilities, instead. This approach allows users to interact however they want, and it is up to the system to recognize whether it needs to change and do so accordingly.

But what about systems that don't have awareness or when an automatic adaptation could produce more barriers or unwanted changes?

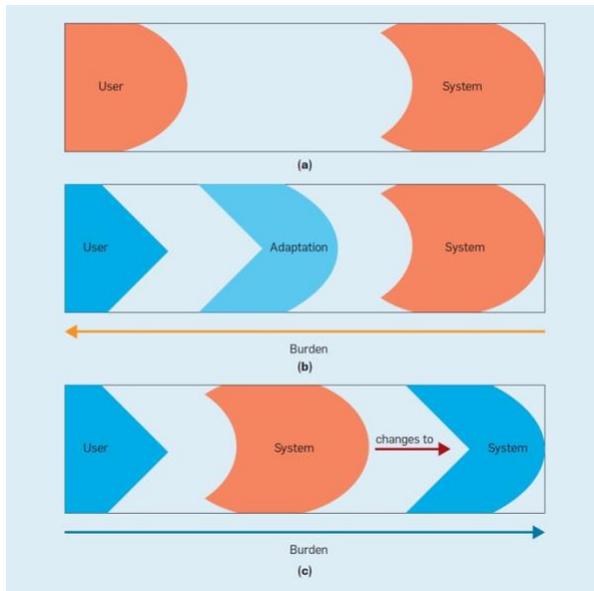

**Fig 2**: (Wobbrock et al's figure [13]) User abilities and a system's ability assumptions: (a) user abilities match a system's ability assumptions; (b) in assistive technology, the user acquires an adaptation to remedy a mismatch; and (c) in ability-based design, user abilities drive changes in the system.

## 2.1 Automatically-adapting systems are not always an appropriate design choice

In cases where one user's fit is another user's barrier (which we call *access friction* [6]), systems must be able to adapt. Ability-based design proposes that systems could have sensory awareness and a degree of decision-making or intelligence. These systems would sense the user, their behavior, or their environment and adapt *automatically* to a perceived interaction barrier.

But many computing contexts lack the hardware capabilities to detect a user's body and interaction patterns. In addition, some contexts (such as the web) are standardized to protect the privacy of assistive technology users and obfuscate or hide which input devices are used through layers of abstraction [8]. Programmatic detection of the user's body and abilities is seen as a potential privacy and security issue for many reasons. To further this problem space, some adaptations a system makes automatically may even produce additional or unwanted barriers or features without the user's consent.

In addition, there are also important social and contextual factors that could influence the needs of a user or users who are interacting with a technology that simply cannot or should not be part of a system's design considerations. In particular, there may be cases where users are collaboratively or *interdependently* interacting with a technology and their social dynamic is the place where access friction is negotiated, not the system [1]. Even in perfect hardware and software conditions, automatic adaptation may not be produce sufficient or ideal outcomes.

## 2.2 Options should not compensate for inaccessible design

The rise of accessibility options in video games in particular in the past 2 years has led to debates on social media about whether options should be seen as award-worthy design on their own or not [10]. I am writing in agreement with what Grant Stoner, Morgan Baker, Mila Pavlin, Ian Hamilton (and others) have already written about in terms of games to remark on broader computing contexts as well: While accessible-by-design is not sufficient by itself, options should not be used to compensate for a design that was inaccessible by default. The perspective that these accessibility folks from the gaming industry have about games is also applicable to many other computing contexts.

ODD has strong advantages when accessible designs for one user may end up producing cognitive, functional, or presentational barriers for another user. However, expecting users to navigate access friction between a design's default settings and their needs and preferences on their own behalf has limitations.

In systems that are inaccessible, burden is placed on the user to adapt to the system's ability assumptions. In ability-based design, the burden is placed on the system to adapt to the user. But where is the burden in ODD?

Unfortunately, the burden is bi-directional in option-driven design (see Fig 1). In an ODD approach, the system must still be capable of change. This means that a system must be designed to be technically robust, allowing for different inputs, outputs, logic, flow, processes, and more. However, assuming users know which accessibility settings they need, how to find them, and how given options help address the barriers that exist in a system is a significant burden.

In an ODD approach, users must be able to know, either during the use of the system or before using it, that something about the system is not only insufficient but also *can* be changed. In my own professional work, I often find that most users not only assume that systems cannot be changed but are already so used to having to adapt to systems that they are not willing to find and manipulate system settings unless they intend to use the system long enough or already know that a given system can be adjusted.

Option-driven design often has findability and understandability burdens to overcome, in addition to costing a user time and patience.

# 3. Design Considerations & Tradeoffs
The following section engages important tradeoffs for designers to weigh when considering an *options-driven design* approach.

## 3.1 Time-of-use is a key variable
How long will a user interact with a system? Just one hour total, like a webpage? One hour every day, perhaps like an email interface? A few hours a day, like an internet browser? Or constantly, like a desktop or mobile operating system? The time a user is expected to spend occupying a given system is one of the most important considerations when choosing whether ODD is an appropriate approach.

*3.1.1 Long-use contexts*
When the context is a mobile or desktop operating system, the decision to have a few options provided has more worth to the time investment of a user. This is especially true if all applications and additional software within that system also inherit the same settings (which presently is enforced better in some ecosystems, to put it lightly).

Applications that might occupy a significant and regular amount of time for a user, such as a social media app or an app used for the sake of income or employment, the benefit of adjusting options may be worth the user's time as well.

*3.1.2 Medium-use contexts*
Video games, currently an industry that commands more revenue than all of film and music combined [12], occupies a middle ground for ODD when considering time-of-use. A triple-A (AAA) game may expect to occupy a player's time from anywhere between 10 and 400 hours (or more). Despite this wide range of use, some award-winning AAA games have more than 60 accessibility settings [9] while others which are still lauded for their accessibility by their players have literally none at all [11]. All games benefit from being accessible by design, but no two games offer the same value to a player for the same option.

Another important consideration for games specifically is that options navigate an essential tension in games between accessibility and difficulty. Ian Hamilton explains that because games hinge on overcoming challenges, any options at all interact with the challenges the user experiences [3]. So for games, finding that ideal experience that doesn't block users from playing but still provides a sense of challenge is key to the design process. Other contexts by comparison, such as websites or operating systems, should not strive to be difficult at all.

*3.1.3 Short-use contexts*
On the other end of the spectrum of ODD and time-of-use are websites. Websites are an example of a context where ODD is not only rarely worth a user's time, but also goes against the philosophy of web accessibility (which expects accessibility by default) [4].

Overlay solutions (which are not only riddled with lawsuits and technical barriers [2]) largely rely on an ODD approach: users are expected to adjust accessibility settings (such as contrast, text size, and animations) on their own every time they visit a new webpage that uses an overlay. In addition to this, overlays don't share settings for the same user across sites that use the same overlay, don't inherit higher level settings (for large font or high contrast) from the operating system or browser, and have no standard for sharing their settings with other overlay venders.

Largely, ODD has become a time-consuming expense for users with disabilities on the web and primarily absolves website owners and maintainers from pursuing an accessible-by-design approach in the first place (it is often the selling point of an overlay that it can "make your website accessible" with a single line of code).

## 3.2 Modularity & extensibility are also options
The customizability and modular nature of some games or software, such Visual Studio Code or Bethesda's Fallout and Skyrim, are closely related to option-driven design. With ability assumptions, a designer does not expect user adaptations. With ability-based design, the design expects specific adaptations and provides a way for the system to adapt automatically. With option-driven design, the designer expects specific adaptations and provides a means for the *user* to enact those on the system. Modularity and extensibility is a type of option-driven design, except that the designer does not expect specific adaptations but instead provides a means for users themselves to identify and adapt the system. In addition, modular systems also often provide a way for users to share their adaptations with others. In this sense, the burden in this model is placed on the system, users, *and* community members and infrastructure.

But for users who aren't community contributors, their experience of the options available to them are similar to designer-curated options, except that the documentation, functionality, and maintenance of a given option are determined by the community instead or the system's dedicated or core designers.

VS Code as a software is built with reasonable core defaults (as well as some pre-determined options) but can also be almost wholly customized through community-contributed extensions (and is designed to encourage users to do this).

However, it is essential to note that extensions and mods built by community members that are used for accessibility purposes, such as Chrome's Dark Reader, pose important questions about who should be responsible for the accessibility of a system, whether it is ethical for software designers to rely on community-*designed* extensions, and whether software builders should rely on community-*maintained* extensions.

As an example, World of Warcraft's most competitive and difficult content has arguably been intentionally designed around the expectation that players will continue to be able to use a mod that is built and maintained by a single person [7]. It may not be ethical for software, after recognizing their access barriers, to expect that the community continues to maintain ways to navigate those barriers.

### *3.3 Inheritable options can take some of the burden away from the user*

In closed ecosystems, like Apple's for example, accessibility settings propagate between systems and are inherited by centralized, set-once-and-forget accessibility options. The iPhone alone has dozens of accessibility settings that influence the entire interaction design of applications. This is an ideal example.

However, generally systems that scale struggle to provide solutions that can also accommodate and fit user's needs and preferences [5]. Many ecosystems lack standards for interoperability. Video games in particular do not have persistence in settings across new games. Players must continually search for and set their options again with every new game they play, and many games do not offer a standard or similar set of options. In addition, since accessibility overlays on websites are intended to serve the design of the website by default and not the user, must have contrast, text size, and other options set every time a user visits a new site. Overlay vendors do not share a user's settings with one another (and should not for the sake of their privacy) but also don't inherit higher levels of settings that a user specifies, such as using Windows High Contrast mode.

But with the apparent demand for ODD's advantages in ideal settings, helping users navigate access friction and contexts where their needs cannot and should not be known, it is important for technical standards for interoperability to arise within specific domains. Video games may not benefit from standardized elements and semantics at an individual game level, but likely could benefit from standardization at the level of a console or operating system. Websites also are a context where there should be more standardization to inherit both browser-level and operating system-level user settings.

By working towards inheritability and solutions that can create more fluid adaptations, we can retain many of the advantages of an ODD approach while also still relieving the burdens placed on users.

## 4. Discussion & Suggestions

Option-driven design as a strategy for accessibility has been sorely under-discussed in academic circles, especially. And with the rise of both accessibility in games and overlays for websites, knowing when to suggest ODD and when to avoid it is still a murky space for designers. I hope that this micro-paper primarily serves to do a few things:

1. Invigorate research attention to look at how users with disabilities interact with systems in different contexts to get a better sense of when ODD is an appropriate choice, despite the burden placed on a user.
2. Stimulate a technical conversation around standardizing *inheritable* options in ecosystems like game consoles, so that users don't have to re-specify the same settings with every new game.
3. Push designers and the general public to critically engage contexts where ODD is used to enable bad design practices, such as overlays on websites or in applications that are not accessible by default.
4. Imagine new paradigms for design that allow more users more persistent control and dynamic expressiveness over their experiences. (In what contexts might users want to be their *own* designers and how can we shape technology to serve their goals?)

## 5. Conclusion

There is a sweet spot for option-driven design that must be carefully considered. Everyone, from researchers, engineers, designers, accessibility practitioners, to players and users, should be able to recognize the tradeoffs and considerations of option-driven design. As a nearly ubiquitous interaction design pattern in computing, it has risen in popularity to both solve legitimate problems posed by complex interactive systems and as a band-aid that enables poor design practices to continue.

It is important to look into the future and treat option-driven design as just one strategy among many that can be employed when engaging accessibility. It is, after all, just an *option*. And it is one that should be considered carefully.

## Acknowledgements

I just want to give enormous thanks to Jonathan Zong for being an ardent supporter of my musings. Thanks also to Shuli Jones, Yunzhi Li, Joon Jang, Sanika Moharana, and Franklin Li for feedback and discussion.

# References

[1] Cynthia L. Bennett, Erin Brady, and Stacy M. Branham. 2018. *Interdependence as a Frame for Assistive Technology Research and Design.* In Proceedings of the 20th International ACM SIGACCESS Conference on Computers and Accessibility (ASSETS '18). Association for Computing Machinery, New York, NY, USA, 161–173. https://doi.org/10.1145/3234695.3236348

[2] Karl Groves. 2021. Overlay fact sheet, Overlay Fact Sheet. Available at: https://overlayfactsheet.com/ (Accessed: April 18, 2023).

[3] Ian Hamilton. 2021. *Difficulty vs accessibility*, YouTube. YouTube. Available at: https://www.youtube.com/watch?v=sPehhHZvKE8 Accessed: April 18, 2023.

[4] Ian Hamilton. 2022. "Websites don't need accessibility options, that's not how websites work." *A thread on Twitter by @ianhamilton_,* Accessed: March 4, 2023. https://twitter.com/ianhamilton_/status/1631763159508828162

[5] D.L. Hickman and D.A. Hagerty. 2021. *Standardised access: The tension between scale and fit.* Ada Lovelace Institute. Available at: https://www.adalovelaceinstitute.org/blog/standardised-access-tension-scale-fit/ Accessed: April 18, 2023.

[6] Travis Chi Wing Lau. 2023. "Access friction can be deeply painful: to feel as though your needs may in fact conflict with another's. That your needs, that which sustains you, may make conditions less accessible for another.." *A thread on Twitter by @travisclau,* Accessed: April 18, 2023. https://twitter.com/travisclau/status/1646604032541184001?s=20

[7] Cass Marshall. 2018. World of warcraft community rallies for the creator of a beloved mod, Polygon. Polygon. Available at: https://www.polygon.com/2018/9/25/17901552/world-of-warcraft-deadly-boss-mods-patreon Accessed: April 18, 2023.

[8] Y. -S. Martín, J. M. del Alamo and J. C. Yelmo. 2014. *Engineering privacy requirements valuable lessons from another realm.* 2014 IEEE 1st International Workshop on Evolving Security and Privacy Requirements Engineering (ESPRE), Karlskrona, Sweden, pp. 19-24, https://doi.org/10.1109/ESPRE.2014.6890523.

[9] Playstation. *The last of us part II - accessibility*, PlayStation. Available at: https://www.playstation.com/en-us/games/the-last-of-us-part-ii/accessibility/ Accessed: April 18, 2023.

[10] Grant Stoner. "When we think of accessibility, the first response is to always fight for OPTIONS, OPTIONS, OPTIONS." *A thread on Twitter by @Super_Crip1994,* Accessed: March 4, 2023. https://twitter.com/Super_Crip1994/status/1463265279060889603

[11] Grant Stoner. "The mainline Pokémon games are INCREDIBLY accessible for physically disabled players, and they really don't feature any form of accessible options." *A thread on Twitter by @Super_Crip1994,* Accessed: March 4, 2023. https://twitter.com/Super_Crip1994/status/1463265281095180290

[12] Valentine, R. (2021) *Digital Games spending reached $127 billion in 2020*, *GamesIndustry.biz*. SuperData. Available at: https://www.gamesindustry.biz/digital-games-spending-reached-usd127-billion-in-2020 (Accessed: April 29, 2023).

[13] Jacob O. Wobbrock, Krzysztof Z. Gajos, Shaun K. Kane, and Gregg C. Vanderheiden. 2018. *Ability-Based Design.* Communications of the ACM. (May 2018), 9 pages. https://doi.org/10.1145/3148051

[14] Jacob O. Wobbrock, Shaun K. Kane, Krzysztof Z. Gajos, Susumu Harada, and Jon Froehlich. 2011. *Ability-Based Design: Concept, Principles and Examples.* ACM Trans. Access. Comput. 3, 3, Article 9 (April 2011), 27 pages. https://doi.org/10.1145/1952383.1952384